\documentclass[preprint,showpacs,preprintnumbers,amsmath,amssymb]{revtex4}

\usepackage{graphicx}
\usepackage{dcolumn}
\usepackage{bm}

\begin{document}

\preprint{}

\title{Vector Polarons in a Degenerate Electron System}

\author{Dennis P. Clougherty}
\email{dpc@physics.uvm.edu}
\author{Charles A. Foell}
\email{cfoell@uvm.edu}
\affiliation{
Department of Physics\\
University of Vermont\\
Burlington, VT 05405-0125}

\date{\today}

\begin{abstract}
We consider a one dimensional model of an electron in a doubly (or nearly) degenerate band that interacts with elastic distortions.  We show that the electron equations of motion reduce to a set of coupled non-linear Schr\"odinger equations.  For the case of interband electron-phonon coupling stemming from local Jahn-Teller interactions, new multicomponent self-localized polaron solutions --vector polarons-- are described and classified. The phase diagram for the different types of vector polarons in this model is presented.  By interpreting the components of the orbital doublet as those of spin-$\frac{1}{2}$, our results can also be used to describe bound magnetic polarons.
\end{abstract}


\maketitle


Nonlinear excitations are known to play an important role
in shaping the optical and transport properties \cite{ssh, bishop83} in quasi-one-dimensional electron systems.  The best-studied example is the kink soliton in trans-polyacetylene, as elucidated in the seminal work of Su, Schrieffer, and Heeger \cite{ssh}.  Many studies have concluded that there are a number of technologically intriguing materials whose properties are determined by solitons, polarons and bipolarons such as conducting polymers \cite{ssh, bishop83, heeger}, organic \cite{lin} and high-Tc superconductors \cite{mott}, halogen-bridged mixed valence transition metal (MX) chains \cite{mx1, mx2}, and the manganites \cite{dagotto, billinge}.  Recently discovered nanoscale materials such as polymeric fullerides and carbon nanotubes may well be added to this list.

We consider the effects of local Jahn-Teller ($E\otimes b_1$) and pseudo-Jahn-Teller ($(A_1\oplus B_1)\otimes b_1$) electron-phonon coupling in a continuum model.  Such a model might apply to a one-dimensional molecular crystal with a two-fold electronic degeneracy or near degeneracy.  The existence of a double degeneracy is insured with sufficiently high spatial symmetry such as rotational symmetry axis three-fold or higher.  The simplicity of the model considered here results in an analytic treatment of the wavefunctions, binding energies, and phase diagram.  

It was proposed long ago that Jahn-Teller coupling plays an important  role in both high-Tc 
superconductivity \cite{johnson1} and fullerene superconductivity \cite{johnson2}.  More recently Jahn-Teller polarons have found application in the manganites \cite{dagotto, billinge}.  Thus polarons of the kind described here has relevance to several systems of current interest.

On the basis of our model, we predict a new class of polarons in degenerate (or nearly degenerate) electron systems that couple to a bimodal distortion.
We find that the model admits three types of stable vector soliton solutions.  These soliton states consist of a localized electron with concomitant distortion of the elastic continuum -- a polaron -- having a spatial structure that involves a mixing of the degenerate (or nearly degenerate) orbital states.   The electronic wavefunction of the solution is the direct analog of the envelope of the $\vec E$-field in the vector solitons studied in non-linear birefringent optical fibers \cite{malomed} or the condensate wavefunction in spinor Bose-Einstein condensation \cite{ho}.

We observe that our study has applications to magnetic materials; one can reinterpret the two orbital components as those of a two-component spinor and apply our results to systems supporting magnetic polarons.


{\it The Model:}  We consider a single electron interacting with local, classical, bimodal distortions of the continuum.
We take the Lagrangian to be the following
\begin{equation}
{\cal L}={\cal L}_e+{\cal L}_{ph}+{\cal L}_{ep}
\end{equation}
with
\begin{subequations}
\begin{equation}
{\cal L}_e={i \over
2}(\Psi^\dagger\partial_t\Psi-\Psi^T\partial_t\Psi^*)-
{1\over 2}\partial_x\Psi^\dagger\partial_x\Psi-\Delta \Psi^\dagger \tau_3\Psi
\end{equation}
\begin{equation}
{\cal L}_{ph}=-\sum_i{\kappa_i\over 2}\phi_i^2
\end{equation}
\begin{equation}
{\cal L}_{ep}= g_0\phi_0\Psi^\dagger \Psi +g_1\phi_1\Psi^\dagger \tau_3\Psi
\end{equation}
\label{lagrange}
\end{subequations}
$\tau_3$ is the usual Pauli matrix.

We have used a two-component representation for our orbital basis with
\begin{equation}
\Psi=\left(\begin{array}{c}
\psi_1\\
\psi_2\\
\end{array}
\right)
\end{equation}
${\cal L}_e$ describes a doublet band of free carriers with splitting $2\Delta$.  ${\cal L}_{ph}$ describes two local classical distortions of the continuum with elastic constants $\kappa_0$ and $\kappa_1$.  We neglect the kinetic energy of the distortion, consistent with the adiabatic approximation.  ${\cal L}_{ep}$ gives the linear coupling of the free carriers to the local distortions: $\phi_0$ preserves the local spatial symmetry, as a breathing mode would; $\phi_1$ is a Jahn-Teller active mode that breaks the local symmetry and mixes the electronic states.  We will refer to $\phi_0$ as the Holstein mode, as it couples to the total electron density in the usual way of the Holstein model \cite{noack}.  Atomic units are used where $\hbar=m=1$.

The following equations of motion result from this Lagrangian:
\begin{subequations}
\begin{equation}
{\phi_0}={g_0\over\kappa_0}(\Psi^\dagger \Psi)
\end{equation}
\begin{equation}
{\phi_1}={g_1\over\kappa_1}(\Psi^\dagger \tau_3\Psi)
\end{equation}
\label{phonon}
\end{subequations}

We use Eq.~\ref{phonon} to eliminate distortions from the remaining equations of motion, yielding

\begin{subequations}
\begin{equation}
i\partial_t\psi_1-\Delta \psi_1 +{1\over
2}\partial^2_x \psi_1+\big( \nu |\psi_1|^2+ \eta |\psi_2|^2\big)
\psi_1=0
\end{equation}
\begin{equation}
i\partial_t\psi_2+\Delta \psi_2 +{1\over
2}\partial^2_x \psi_2+\big( \nu |\psi_2|^2+\eta |\psi_1|^2\big)
\psi_2=0
\end{equation}
\label{bg1}
\end{subequations}

with
\begin{subequations}
\begin{equation}
\nu={g_0^2\over\kappa_0}+{g_1^2\over\kappa_1}\equiv \gamma_0^2+\gamma_1^2
\end{equation}
\begin{equation}
\eta={g_0^2\over\kappa_0}-{g_1^2\over\kappa_1}\equiv \gamma_0^2-\gamma_1^2
\end{equation}
\label{gammas}
\end{subequations}
We observe that $\nu$ is non-negative, while $\eta$ can change sign depending on the relative importance of the coupling to the Jahn-Teller mode $\phi_1$ compared to the Holstein mode $\phi_0$.  In the context of the analogous system of equations in non-linear optics, $\nu$ is referred to as the self-modulation coefficient, while $\eta$ is the cross-modulation coefficient.

Consistent with Galilean invariance, we consider time-dependent solutions of the following form:

\begin{eqnarray}
\psi_j=\sqrt{ |E-\Delta | \over  \nu }\  {\rm exp}\bigg[{i\bigg({ v x}-(E+{1
\over 2}  v^2)t\bigg)}\bigg]\cr
\times \ r_j\bigg({\sqrt{2  |E-\Delta |}}\ 
(x-v t)\bigg)
\end{eqnarray}

A set of coupled nonlinear Schr\"odinger equations for the envelope functions $r_j$ results
\begin{subequations}
\begin{equation}
r_1''(\xi)+{E-\Delta \over |E-\Delta|} r_1(\xi)+r_1(\xi)\bigg(
{r_1^2(\xi)}+{\eta\over \nu } {r_2^2(\xi)}\bigg)= 0
\end{equation}
\begin{equation}
r_2''(\xi)+{E+\Delta \over |E-\Delta|} r_2(\xi)+r_2(\xi)\bigg(
{r_2^2(\xi)}+{\eta\over \nu } {r_1^2(\xi)}\bigg)=0
\end{equation}
\end{subequations}

Self-localized solutions exist only for $E < 0$ and $|E|> \Delta$. In addition, known
localized, stable solutions only exist for $\nu$ and $\eta$ non-negative \cite{pelinovsky}.

\begin{figure}
\includegraphics[width=3in]{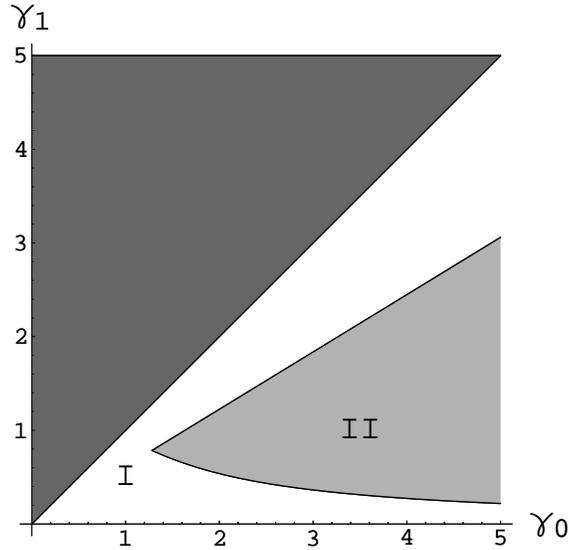}
\caption{\label{pd} Plot of the polaron phase diagram.    Type I polarons are stable in region I, and type II polarons are stable in region II.   Type III polarons stably exist for $\gamma_1=0$ (horizontal semi-axis). (We take the arbitrary case of $\Delta=0.1$.)}
\end{figure}

We define $\omega ^2 \equiv
(|E|-\Delta)/(|E|+\Delta)$ and $\beta \equiv \eta/\nu$, giving

\begin{subequations}
\begin{equation}
r_1''-r_1+r_1\bigg({r_1}^2+\beta {r_2}^2\bigg)=0
\end{equation}
\begin{equation}
r_2''-\omega ^2 r_2+r_2\bigg({r_2}^2+\beta {r_1}^2\bigg)=0
\end{equation}
\label{se}
\end{subequations}
with $0\le \beta \le 1$ corresponding to the condition for known stable polaron solutions (see Fig.~\ref{pd}).


{\it Polaron Types:}  We describe three distinct types of stable polaron solutions to Eqs.~\ref{se}:  type I  has equal amplitudes $r_1(\xi)=r_2(\xi)$, corresponding to an equal orbital mixing that is constant over the spatial extent of the polaron.  It results when $\omega=1$ and $0\le\beta < 1$.  With $\omega=1$, the uncoupled electronic states must be exactly degenerate ($\Delta=0$).  

\begin{equation}
r_1(\xi)= r_2(\xi) = \sqrt{2\over 1+\beta}\  {\rm sech}\ \xi
\end{equation}

The {type II} polaron has unequal amplitudes and results when $\omega\ne 1$. The value of $\beta$ is determined by $\omega$.   The uncoupled electronic states are nearly degenerate. 

\begin{figure}
\includegraphics[width=3in]{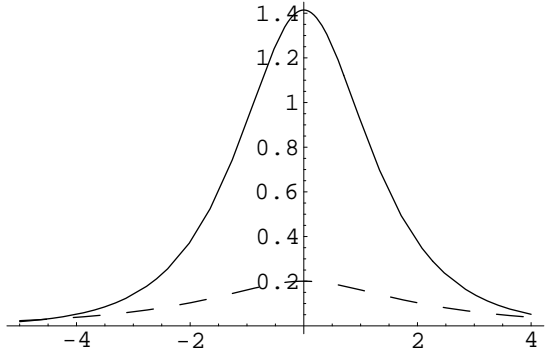}
\caption{\label{wdw} Plot of type II polaron ($\alpha=0.2$ and $\omega=0.6$). Solid line is $r_1(\xi)$, and $r_2(\xi)$ is dashed.}
\end{figure}

In the wave-daughter wave approximation, we take
$|r_1|\gg |r_2|$.   To leading order in the daughter wave amplitude, the following solution was determined by Yang and Benney \cite{yang96, yang97}.

\begin{subequations}
\begin{equation}
r_1(\xi)=\sqrt{2}\  {\rm sech}\ \xi
\end{equation}
\begin{equation}
r_2(\xi)=\alpha\  {\rm sech}^\omega \ \xi
\end{equation}
\end{subequations}
where $\alpha\ll 1$ and 
\begin{equation}
\omega =  {1\over 2} ( \sqrt{1+8 \beta} -1)
\label{omega}
\end{equation}

The type II polaron is unbound unless the coupling strengths satisfy

\begin{equation}
\gamma_0^2+\gamma_1^2 > \sqrt{24\Delta}
\label{b1}
\end{equation}
Just as in the case of the impurity pseudo-Jahn-Teller effect, we observe that there is a critical coupling strength that must be exceeded for a distortion to develop \cite{dpc98}.

In addition, Eq.~\ref{omega} and the binding energy relation given in Table  \ref{mass} lead to the constraint that type II solutions exist only when
\begin{equation}
\beta \ge {1+\sqrt{3}\over 6}
\label{b2}
\end{equation}
This constraint consequently limits the range of $\omega$ such that $(\sqrt{21+12\sqrt{3}} -3)/6 \le  \omega  <1$.

The inequality of Eq.~\ref{b2} and the binding energies of Table \ref{mass} lead to the boundaries of the region of stability for type II polarons in the model's parameter space.  Equating the binding energies for type I and type II polarons gives the coexistence curve

\begin{equation}
\gamma_1^2=\sqrt{\gamma_0^4+24\Delta}-\gamma_0^2
\end{equation}

The {type III} polaron results in the case of $\beta=1$ and $\omega=1$.  The equations in this case were studied first by Manakov \cite{manakov} in another context.  The condition of $\beta=1$ implies that $\gamma_1$ must vanish.  The special symmetry that results gives an infinite set of degenerate solutions, labeled by a continuous parameter $0\le\theta< 2\pi$. This continuous internal degree of freedom allows one to qualitatively distinguish between the $E\otimes b_1$ polaron and the Holstein case.

\begin{subequations}
\begin{equation}
r_1(\xi)=\sqrt{2} \cos\theta\  {\rm sech}\  \xi
\end{equation}
\begin{equation}
r_2(\xi)=\sqrt{2} \sin\theta\  {\rm sech}\ \xi
\end{equation}
\end{subequations}

\begin{figure}
\includegraphics[width=3in]{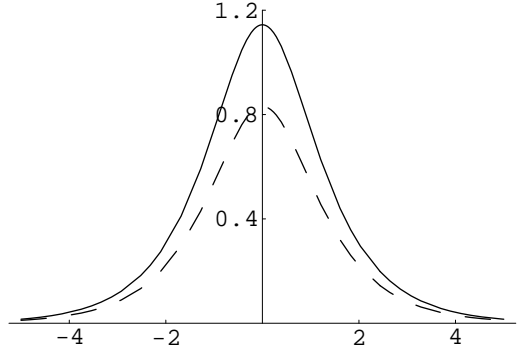}
\caption{\label{sol1} Plot of type III polaron ($\beta=1$ and $\omega=1$) for $\theta={\pi\over 5}$. Solid line is $r_1(\xi)$, and $r_2(\xi)$ is dashed.}
\end{figure}

\begin{table}
\begin{ruledtabular}
\begin{center}
\begin{tabular}{cc}
\hline

Polaron type &  Binding Energy \\

\hline {Type I }& 

${1 \over 24}\gamma_0^4$ \\

\hline Type II &

$ \bigg({(\gamma_0^2+\gamma_1^2)^2 \over 24}-\Delta\bigg)$ \\

\hline Type III &

${1 \over 24}\gamma_0^4$ \\

\hline
\end{tabular}
\caption{\small Polaron binding energies. (Type II binding energy is given to ${\cal O}(\alpha)$.)} 
\label{mass}
\end{center}
\end{ruledtabular}
\end{table}

Fig.~\ref{pd} gives the phase diagram for the three types of vector polarons.  The upper region corresponding to $\gamma_1 > \gamma_0$ does not contain bound polarons.  In the region $\gamma_1 \le \gamma_0$, type I polarons are stable throughout; however, type II polarons have a lower binding energy in the wedge-shaped region (light gray).  The type I polarons continue onto the locus of points $\gamma_1=0$, becoming type III polarons.

The one dimensional model studied has three types of stable two-component soliton solutions.  These nonlinear excitations correspond to bimodal (pseudo) Jahn-Teller polarons, where degenerate (or nearly degenerate) orbital states are mixed under coupling to a localized distortion of the continuum.  Such states might be produced in quasi-one dimensional systems that are lightly doped by  charge-transfer impurities.  

Type I and type III solutions have two important features in common: (1) the orbital mixing is constant in space, and (2) the wavefunction decay far from the polaron center has the same asymptotic form, going as ${\exp}(-|\xi|)$. We observe that type II solutions have a position-dependent mixing that results from the two components having different asymptotic forms, with the primary wavefunction going as ${\exp}(-|\xi|)$ and the daughter wavefunction behaving as ${\exp}(-\omega|\xi|)$.  The amplitudes of the correlated distortions fall off twice as fast with distance.

Finally we note that by interpreting the two states as the components of spin-$\frac{1}{2}$  rather than those of an orbital doublet, the Lagrangian given in Eq.~\ref{lagrange} also describes an electron interacting with local {\it magnetization} $\phi_1$.  Thus these polarons can be thought of as new kinds of bound {\it magnetic} polarons in a system with an internal effective magnetic field of $\Delta$ which could be the result of exchange splitting or an externally applied field.  Such states might have application to transport in doped magnetic semiconductors and other systems where there is interplay between electron-phonon coupling and magnetic exchange coupling.

\end{document}